\documentclass[conference]{IEEEtran}
\usepackage{cite}

\ifCLASSINFOpdf
  \usepackage[pdftex]{graphicx}
\else
  \usepackage[dvips]{graphicx}
\fi
\usepackage{url}

\usepackage{multicol}
\usepackage{color}
\usepackage{microtype}
\usepackage{paralist} 

\begin{document}
%
\title{VisIt: Experiences with Sustainable Software}


\author{\IEEEauthorblockN{Sean Ahern}
\IEEEauthorblockA{Oak Ridge Nat'l Lab.\\
ahern@ornl.gov}
\and
\IEEEauthorblockN{Eric Brugger}
\IEEEauthorblockA{Lawrence Livermore Nat'l Lab.\\
brugger1@llnl.gov}
\and
\IEEEauthorblockN{Brad Whitlock}
\IEEEauthorblockA{Intelligent Light, Inc.\\
bjw@ilight.com}
\and
\IEEEauthorblockN{Jeremy S. Meredith}
\IEEEauthorblockA{Oak Ridge Nat'l Lab.\\
jsmeredith@ornl.gov}
\and
\IEEEauthorblockN{Kathleen Biagas}
\IEEEauthorblockA{Lawrence Livermore Nat'l Lab.\\
biagas2@llnl.gov}
\and
\IEEEauthorblockN{Mark C. Miller}
\IEEEauthorblockA{Lawrence Livermore Nat'l Lab.\\
markmiller@llnl.gov}
\and
\IEEEauthorblockN{Hank Childs}
\IEEEauthorblockA{Lawrence Berkeley Nat'l Lab. \&\\
The University of Oregon\\
hchilds@lbl.gov, hank@uoregon.edu}
}


%


\maketitle


%
\IEEEpeerreviewmaketitle

\section{Introduction}
\label{sec:intro}


Visualization tools have long been fundamental to the process of scientific
discovery. Nowhere is this more true than in the field of high-performance
simulation, where data sets now run into the tens of petabytes in size.  As
the growth of simulation data has exploded in the last decade, so too has
the need for scalable and nimble tools to provide insight into these
complex results.  At the same time that virtually all fields of science
have seen data grow in size and complexity, the computing systems upon
which simulations are run and data analyses are performed have become
similarly complex.  Parallel computing has become the dominant method by
which scientific simulations are done, and visualization infrastructures
have had to embrace this same method to continue to provide capabilities
for scientific understanding.


VisIt was developed in response to these emerging needs.  It is
an open-source project for visualizing and analyzing extremely large data
sets, while still exploiting the graphics capabilities of users' desktops.
The project has evolved around three focal points: (1) enabling data
understanding, (2) scalable support for extremely large data, and (3)
providing a robust and usable product for end users and researchers.

In turn, these focal points have made VisIt a very popular tool for
visualizing and analyzing the data sets generated on the world's largest
supercomputers.  VisIt received an R\&D100 Award in 2005 for the tool's
capabilities in understanding large data sets, it has been downloaded
hundreds of thousands of times, and it is used all over the world.

VisIt's success has been wholly dependent upon the culture and practices of
software development that have fostered its welcome by users and embrace
by developers and researchers.  In the following paper, we, the founding
developers and designers of VisIt, summarize some of the major efforts,
both successful and unsuccessful, that we have undertaken in the last
thirteen years to foster community, encourage research, create a sustainable
open-source development model, measure impact, and support production
software.  We also provide commentary about the career paths that our
development work has engendered.


\section{Funding and Adoption}
\label{sec:history}




Thirteen years ago there was a fundamental change in the way scientific
simulation was being used, where ever increasing numbers of users were
running simulations generating large amounts of data.  This was especially
true within the United States' Stockpile Stewardship program, and the
heroic computational effort known as Advanced Simulation and Computing
(ASC).  It quickly became apparent the existing tools were not scaling to
the size of newer data sets, and it was no longer feasible to have
visualization experts develop and utilize a host of specialized tools to
analyze users' data for them.  What was needed was a robust, flexible, and
general-purpose tool with which end users could analyze and visualize
their data.  A group of visualization developers were able to convince
management at the Lawrence Livermore National Laboratory (LLNL) that what
was needed was a new open-source tool, itself built with open-source
software, that would enable data understanding while providing a foundation
for implementing future scalable algorithms.  Included in this mission was
a mandate to provide a robust and usable product for end users.  VisIt was
born.

A small group within LLNL worked for the next several years
to develop and polish this new tool.  Using an Agile 
development model, the team released new versions frequently while working closely
with users.  This resulted in a popular tool for visualizing and analyzing
data sets on the world's largest supercomputers and culminated in an
R\&D100 award in 2005.

External collaboration initially came from Sandia and Los Alamos National Laboratories.
As adoption grew in the wider scientific community, VisIt
received funding from the Department of Energy (DOE) Scientific Discovery
through Advanced Computing (SciDAC) program as part of the Visualization and
Analytics Center for Enabling Technologies (VACET).
To enable visualization researchers and developers from
multiple institutions to contribute efficiently, development
was transitioned from a LLNL-internal model to a distributed model.
This led to a period of growth in functionality where new
state-of-the-art visualization and analysis techniques were added to the
tool from institutions that were leaders in the areas of visualization and
analysis, such as Lawrence Berkeley National Laboratory, Oak Ridge National
Laboratory, The University of California Davis, and the University of
Utah.

This led to an even wider adoption of VisIt and commercial interest from
companies such as Intelligent Light, Tech-X, and Allinea.  These companies
would either use VisIt to enhance their existing products or create new
products using VisIt.  They also contributed to the user support,
answering questions on the mailing lists.

\vspace{-0.05in}
\section{Fostering Community}
\label{sec:community}
\vspace{-0.05in}

VisIt is a general-purpose visualization tool, and it has been used
effectively in myriad application domains:
climate,
astrophysics,
turbulence,
thermal hydraulics,
engineering,
computational fluid dynamics,
medical, and
many more.
Thus, it has a large potential customer base compared to other HPC
applications.
As a result,
more features and improved quality of implementation
can lead to increases in ``market share,'' which can in turn lead 
to more funding opportunities.

This observation was taken to heart by VisIt developers, perhaps
subconsciously.
After gaining adoption in the ASC program, VisIt developers decided to make
investments that would encourage adoption from new developers and 
customers.
For funding agencies, the benefit of such investments is that they
encourage investment by other agencies, resulting in either reduced cost
for the same product, or in a superior product for the same cost.

The efforts to foster the customer and developer communities were
different, and they are treated separately here.

\subsection{Customer Community}
\label{sec:usercommunity}

For our project, every developer has played roles spanning from researcher to
software engineer to customer support liaison.
In the early phases of the project, the customer base was small enough that
the primary means of user education was personal coaching from developers
on tool usage.
But, as user adoption increased, this model quickly became infeasible, 
as the proportion of time developers spent doing customer support would 
have grown with each new user.
Making investments that scaled developer expertise to many customers became
a good return on investment.

We discuss four methods for our support: traditional documentation,
web support, education, and interactive support.

\textbf{Traditional documentation.}  
Our project has four main manuals.  
These manuals, by and large, were
written by a single developer who recognized their need.
Because of this, the manuals would regularly fall out of date, eventually attracting
the attention of the user community.
This would ultimately prompt a developer to refresh a given manual,
which happened at the rate of every two to three years per manual.
Our fundamental problem was that we did not effectively establish 
project norms that new capabilities
and changes in the interface mandated corresponding changes in
documentation. This was exacerbated by the fact that we did not move 
rapidly to open standards for collaborative document editing and publication.
Going forward, we need to encourage our developers to feel shared ownership
of the quality of the manuals and to develop processes where documentation
updates happen regularly.

\textbf{Web support.}  
The primary face to our project is our web site.
Surprisingly, googling the word ``visit" has returned our project
page as the \#1 result for several years now.
This project page provides access to the source code, pre-compiled
binaries, and documentation, and is fairly static.
The page is complemented by our \textit{``visitusers.org"} site, which is
an evolving site that provides a wiki and a forum.
\textit{``visitusers.org"} was started with the goal of being a community-based site
where VisIt users would publish content and interact via a forum.
We find that the VisIt community is often more interested with obtaining help than
regularly contributing back to the community.
Once consequence of this behavior is that \textit{``visitusers.org"} has effectively
become a repository for superusers---who
often double as developers---for scripts and techniques that enable
complex analyses.
It also is a place where developers list recipes for doing certain types of
visualization, complementing the more formal manuals.

\textbf{Education.}
We have had two primary education activities: tutorials and classes.

We have conducted well over thirty tutorials, ranging in duration from two
to eight hours, at venues geared towards visualization experts (i.e., the IEEE
Visualization conference), high-performance computing experts (i.e.,
Supercomputing), and customers (i.e., the SciDAC conference).
The tutorials were generally well received, but their role in increasing
tool usage is not clear.
The primary struggles with the tutorials are
two-fold: (i) how to teach material to attendees with varying backgrounds
and varying expertise with VisIt?  and (ii) how to get significant material
across in a short period of time?
The contents of the tutorial has gradually changed over the years.
Originally, the tutorial was paced most appropriately for visualization
experts, and many of the demonstrations were designed to inform the audience
about the capabilities of the tool.
Now, the tutorial has many more beginner activities, and the demonstrations
are designed to allow all attendees to have at least a few successful
experiences with the tool.

The VisIt class is taught in a classroom setting where every student has a
computer with VisIt pre-installed.
Many
classes have been taught in partnership with the DOE and National Science
Foundation supercomputing centers (e.g., the Oak Ridge Leadership Computing
Facility).
In this format, an instructor goes through features one by one, and the
students reproduce the results.
The class has one- and two-day versions and has been very well received.
The classes taught at computing centers appear to have a
correlation to tool adoption.
Unfortunately, in contrast with tutorials that can be easily offered at
conferences, classes are a less traditional format for HPC tools.
As a result, the class is taught less frequently and has generally only
been taught at institutions who wish to provide training for their specific
users.

\textbf{Interactive support.} 
Documentation and training are effective, but we have found that users
often need immediate assistance.
We created two mechanisms for this: mailing lists for the general user
community and priority support for our paying customers.

We feel the ``users'' mailing list has been very effective.
Any user may post, but users can only post if they also subscribe.
Ideally, the users would be able to answer each other's questions.
Unfortunately, this does not often happen, and developers answer many
questions.
This is partly because many of the questions are regarding compilation
errors and crashes, and fewer are regarding effective tool usage.
Nevertheless, we view the users mailing list as a success; it provides access
to experts for all users, and expert support often comes quickly, which
users appreciate.

For customer groups who financially support the program, we provide a
higher level of support.
First, we have a separate email list that goes directly to developers and
gets priority.
Second, we created a phone hotline that allows priority users to speak to
an expert.
We have found that these mechanisms have been well-received by our user
community.
Visualization is an interactive process, and the ability to receive
coaching on demand prevents frustrating experiences.

Finally, as our developer community has grown, we have discovered that many
HPC centers have a staff ``expert'' for VisIt who is able to provide direct
local support for users of that center, then falling back upon the wider
developer community as necessary.
In the future, commercial companies that contribute to VisIt may offer paid
consulting support to the wider VisIt user community.

\subsection{Developer Community}
\label{sec:devcommunity}

Our team employs software engineering practices that enable effective
distributed support.
We use a public Subversion repository located at LBNL for revision control
and follow a model with a branch for releases and a branch for the trunk.
We have a bug tracker and, for many bugs, we schedule the release for when
particular bugs will be fixed.
We run a nightly regression test, with results posted to the web, to catch
any new bugs that might have been introduced.
Overall, these practices appear to be effective.

In VisIt's early years, when all development happened within LLNL, the
software systems that supported the activities of revision control, bug
tracking, and revision testing were all done on internal systems with no
outside access, sometimes on commercial software development platforms.
Though these practices were successful, the need to embrace contributions
from external contributors required us to migrate all of these roles to
publicly-accessible servers and open systems.  Though this process
took several years, it is now complete and has proven valuable in nurturing
our developer community.

The first five years of VisIt's development occurred with all of its
primary developers within one hallway.
This co-location was excellent for maintaining consistency in the code and
very efficient for this team of developers, but it likely stunted developer
documentation.
As external developers joined in, documentation began to appear as
questions arose from the new developers, primarily in the form of wiki
pages on \textit{``visitusers.org."}
There is a mailing list for developers to ask other developers
questions about how to develop code.
This list has been successful, and we have found that all developers
are eager to participate on issues where they have expertise.

We adopted a software plug-in model in the early years of the project and
made the main VisIt code only be aware of the abstractions for rendering
and data manipulation techniques and file format readers.
We believe that new developers succeed most often when working on new
plug-ins (i.e., new derived types of the abstractions for rendering, data
manipulation, or file format readers), likely because they are protected
from the complexities of VisIt's implementation and can focus on their own
self-contained code.
On the other hand, developers who must work on the main code face a
steeper learning code.
In short, the software components we designed to be most easily
extensible have been highly maintainable, but the rest of the code base,
though still well-designed and modular, does not show the same degree of
flexibility and independence.

We have
found that the developer team grows by existing developers hiring new
developers at their own institution and training them.
Often, the project spreads to new institutions from developer
migration, not from new developers at a site picking up the tool and
learning it in isolation.
This may indicate a failure in terms of fostering developer community, but
we note that the goal is hard; learning how to develop a program that
exceeds a million lines of code is difficult even with excellent documentation.

\vspace{-0.05in}
\section{Research and Architecture}
\label{sec:research}
\vspace{-0.05in}

Understanding how visualization research is critical to the success of any
future-looking scientific discovery effort, we architected VisIt to be
amenable to a wide range of research activities.  The plug-in model
mentioned above is also critical to the transition of research into
deployed software.

VisIt uses a ``client/server'' model, where the bulk of the I/O
and computation occurs at the large HPC centers, close to the data, while
the interactive rendering occurs at the user's desktop.  This model allows
for the easy deployment of remote visualization capabilities.

Internally, VisIt uses data-parallel pipelines for task- and
data-independence~\cite{Childs:2005}.  Each ``filter,'' or component, of a
pipeline is by-and-large independent from any other filter, allowing for a
vast array of possible data analysis and visualization activities to be
applied to arbitrary data sets.  This method has proven very popular, being
the basis of a number of visualization tools over the last two decades.
This independence also provides a fertile ground for research to explore a
particular element of data analysis or visualization without having to
implement basic functionality like I/O, rendering, or data model
development.

\vspace{-0.05in}
\section{Governance}
\label{sec:governance}
\vspace{-0.05in}




The VisIt team at LLNL is responsible for ensuring that the software
development practices are followed, that quality standards for the tool are
met, that it passes the nightly regression
suite, for setting the release schedule, and for creating the releases.
Changes to the visualization infrastructure are vetted among VisIt developers
via the visit-developers mailing list to ensure that they adhere to the VisIt
design philosophy.

VisIt is funded by several stable long term DOE funding streams, including
the National Nuclear Security Administration and the Office of Science.  It
also receives shorter-term funding for specific enhancements that are of
benefit to small user communities such as the DOE Office of Nuclear
Energy. VisIt has also seen increasing funding from other
federal agencies like the National Science Foundation.

The VisIt project attempts to be as inclusive as possible and will allow
most types of changes as long as it meets basic software engineering
standards and doesn't negatively affect existing users.  Changes that are
localized to a small portion of VisIt that don't impact other user groups
(such as new database reader, operator, or plot) are readily accepted as
long as the code will compile on all the supported platforms.  More
fundamental changes are first discussed on the visit-developers mailing
list to build consensus about the best way to make the change.  The
software architect is also heavily involved in this process.  New VisIt
developers are paired up with experienced developers in an informal
mentoring program where they can learn about the processes and discuss
changes to the code.

\section{Career Paths}
\label{sec:career}
\vspace{-0.05in}

We believe that the project's developers have benefited from
their participation in the project.
The majority of these developers were hired fresh out of
college, and 
many of them had  opportunities
to rise above the individual contributor status 
within three to five years.
One reason for this success is that there were several multi-institution
grants, which created additional opportunities for co-PI status 
for the development team.
Furthermore, each of the developers acted as the VisIt expert
at their home institution and thus often had
the first opportunity to pursue local visualization-related
activities over remote VisIt developers
who may have been even more qualified for the collaboration.
We note that this contrasts with projects where the majority of 
developers are all at one institution and the lead
developer at that institution attracts the lion's share
of grant opportunities, speaking engagements, etc.
Finally, we note that supporting a visualization tool
like VisIt creates significant opportunity to 
network, as the large majority of the HPC community 
are potential customers.
As a result, developers on projects like VisIt may have
additional opportunities in terms of attending
workshops, grants, and speaking engagements.

%
The discussion above represents our observations
of the positives that assisted in career development.
The obvious negative, for a research-oriented developer,
is the additional overhead of developing high-quality
software.

\section{Usage}
\label{sec:usage}
\vspace{-0.05in}

As is the case with many open-source software packages, direct measures of
usage are difficult.
We rely upon the following mostly indirect metrics:
\begin{compactitem}
\item \emph{Usage statistics at individual sites.}
At LLNL, developers collect information on VisIt startups on
a per-user name basis and see approximately 300 unique
user names launching VisIt every month.
At most sites, however, we do not gather this information.

\item \emph{Downloads.}
We measure the number of downloads of both source code and pre-compiled
binaries.  Our binaries have been downloaded over 200,000 times from more than
70 countries internationally.
However, we cannot distinguish multiple downloads by the same user, nor can we
distinguish downloads from actual use, whether routine or occasional.
Traffic at \textit{``visitusers.org"} averages over 35,000 visits per month,
indicating strong interest from VisIt users.

\item \emph{User inquiries.}
We measure activity on our users email lists as well as the institutions
from which they originate.  There are approximately 400 routine subscribers
generating about 300 emails per month.  Over the history of our lists, we
have had regular communication with participants from over 200 worldwide
institutions.

\item \emph{Citations}.  We measure citations of the definitive VisIt
publications.  For example, according to Google Scholar,
\cite{Childs:2005} has received 128 citations,
\cite{Childs:CGA_2010} has received 35,
\cite{childs2006scalable} has 35,
\cite{rubel2008high} has 31,
 \cite{Fogal:2010:LargeData} has 29,
 \cite{Pugmire:SC09} has 24,
and
\cite{whitlock2011parallel} has 23.

\end{compactitem}


These indirect usage metrics necessarily tell an incomplete story.  
However, the alternative\,---\,gathering information
at every startup of user name and site\,---\,is
impossible for classified environments where VisIt is
used, and where such tracking is possible, it is distasteful to many users.
The developers eschew these tactics in order to encourage adoption, but
the price is a lack of more direct usage metrics.

\section{Conclusion}
\label{sec:conclusion}

VisIt's thirteen years of development have seen a significant amount of
success in deploying a scalable open-source tool that has been fundamental
to scientific discovery for the nation. It has also shown a successful model
for nurturing research and fostering its deployment into production for end
user scientists.  Though there have been stumbling blocks along the way, as
is expected in any long-term effort, we believe that the lessons learned by
the VisIt team can be instructive to software efforts looking to have
similar impact and success.



\bibliographystyle{IEEEtran}
\vspace{-0.05in}
\bibliography{IEEEabrv,wssspe}
%

\end{document}